\documentclass[preprint, prX]{revtex4}
\usepackage[all]{xy}
\usepackage{amssymb, amsmath}
\usepackage{accents}
\begin{document}

\title{Generalized five-dimensional Kepler system, Yang-Coulomb monopole and Hurwitz transformation}
\author{Ian Marquette}
\affiliation{School of Mathematics and Physics, The University of Queensland, Brisbane, QLD 4072, Australia}
\email{i.marquette@uq.edu.au}
\begin{abstract}
The 5D Kepler system possesses many interesting properties. This system is superintegrable and also with a $su(2)$ nonAbelian monopole interaction (Yang-Coulomb monopole). This system is also related to a 8D isotropic harmonic oscillator by a Hurwitz transformation. We introduce a new superintegrable Hamiltonian that consists in a 5D Kepler system with new terms of Smorodinsky-Winternitz type. We obtain the integrals of motion of this systems. They generate a quadratic algebra with structure constants involving the Casimir operator of a $so(4)$ Lie algebra. We also show that this system remains superintegrable with a $su(2)$ nonAbelian monopole (generalized Yang-Coulomb monopole). We study this system using parabolic coordinates and obtain from Hurwitz transformation its dual that is a 8D singular oscillator. This 8D singular oscillator is also a new superintegrable system and multiseparable. We obtained its quadratic algebra that involves two Casimir operators of $so(4)$ Lie algebras. This correspondence is used to obtain algebraically the energy spectrum of the generalized Yang Coulomb monopole.
\end{abstract}
\maketitle

\section{Introduction}
In classical mechanics a Hamiltonian system with Hamiltonian H and integrals of motion $X_{a}$

\begin{equation}
H=\frac{1}{2}g_{ik}p_{i}p_{k}+V(\vec{x},\vec{p}),\quad X_{a}=f_{a}(\vec{x},\vec{p}),\quad a=1,..., n-1 \quad,
\end{equation}

is called completely integrable (or Liouville integrable) if it
allows $n$ integrals of motion (including the Hamiltonian) that are
well defined functions on phase space, are in involution
$\{H,X_{a}\}_{p}=0$, $\{X_{a},X_{b}\}_{p}=0$, $a$,$b=1$,...,$n-1$ and
are functionally independent ($\{,\}_{p}$ is a Poisson bracket). A
system is superintegrable if it is integrable and allows further
integrals of motion $Y_{b}(\vec{x},\vec{p})$, $\{H,Y_{b}\}_{p}=0$,
b=n,n+1,...,n+k that are also well defined functions on phase
space and the integrals$\{H,X_{1},...,X_{n-1},Y_{n},...,Y_{n+k}\}$
are functionally independent. A system is maximally
superintegrable if the set contains $2n-1$ functions and minimally
superintegrable if it contains $n+1$ such integrals. The integrals
$Y_{b}$ are not required to be in involution with
$X_{1}$,...$X_{n-1}$, nor with each other.
The same definitions (i.e. integrability and superintegrability) apply in the context of quantum mechanics but $\{H,X_{a},Y_{b}\}$ are well defined quantum mechanical operators, assumed to form an algebraically independent set.
The most well known superintegrable system is the Kepler-Coulomb potential [1-3]. Its integrals of motion consist of the angular momentum and the Laplace-Runge-Lenz vector whose origin can be traced back 80 years earlier [4-6] than Treatise of celestial mechanics of Laplace [7] with the works of Bernoulli [6] and Hermann [4,5].
 
A systematic classification of classical and quantum superintegrable systems was initiated 40 years ago by Smorodinsky, Winternitz and co-workers for Hamiltonians with quadratic (in momenta) integrals of motion in two-dimensional Euclidean space [8]. Four types of maximally superintegrable systems were obtained with multiseparability [8]. This search was extended in three-dimensional Euclidean space, where five potentials are maximally superintegrable with five integrals of motion. Nine potentials belong to the class of minimally superintegrable systems and are characterized by the existence of four second order integrals of motion [9-11]. One of these nine systems include the generalized Kepler system [12-23]

\begin{equation}
H=\frac{1}{2}P^{2}-\frac{c_{0}}{r}+\frac{c_{1}}{r(r+x_{0})}+\frac{c_{2}}{r(r-x_{0})},
\end{equation}
where $r=\sqrt{x_{1}^{2}+x_{2}^{2}+x_{0}^{2}}$ and $c_{1}$ and $c_{2}$ are non-negative constants. When $c_{1}=c_{2}\neq 0$, this system reduces to the Hartmann potential introduced in 1972 as a model for ring-shaped molecules [24]. As the superintegrability property is preserved for the Kepler system with a monopole interaction (MICZ-Kepler) [25-31], the generalized Kepler system with monopole (generalized MICZ-Kepler system) is also superintegrable [32-36]. This system is also related to another superintegrable systems, a 4D singular osicllator [33,36-38], by a Kustaanheimo-Stiefel transformation [39].
Let us mention that the 2D Kepler system is related to the 2D isotropic harmonic oscillator by a Levi-Civita transformation [40]. This relation that map these two superintegrable systems in classical and quantum mechanics was introduced at the origin in context of the regularization of the classical Kepler problem. A such relation also exists for the 5D Kepler system that can be related to a 8D isotropic harmonic oscillator by Hurwitz transformations [41]. This system and the Yang-Coulomb monopole ( i.e. 5D Kepler system with a $su(2)$ non-Abelian monopole) have attracted a lot of attention [42-51]. Let us mention that the existence of the Levi-Civita and its generalizations is one important property of these systems. The Hurwitz relations were exploited to present a derivation of the energy spectrum of the Yang-Coulomb monopole from the ladder operators of the 8D harmonic oscillator [50]. There is an analog of these relations between 1D harmonic oscillator and Kepler system. These four cases are related to the Euler identity  ($x_{0}^{2}+...+x_{d-1}^{2}= (u_{0}^{2}+...+u_{D-1}^{2})^{2}$) which has a bilinear solution in $u_{i}$ only for the following pairs (D,d): (1,1),(2,2),(4,3) and (8,5) [47].

Quadratic algebras and more generally polynomial algebras have been applied to many quantum systems [22,23,36,52-78]. Such algebraic structure generated by the integrals of motion do not only provide an elegant method to obtain the energy spectrum but also explain the degeneracies of the energy spectrum. Recently [36,70,72], it was pointed out how results obtained in context of 2D superintegrable systems and their quadratic algebras can be applied to three and four-dimensional superintegrable systems. 

The purpose of this paper is to study a generalized 5D Kepler system and moreover to obtain the superintegrable analog with a nonAbelian monopole. The study of systems with nonAbelian monopole with Smorodinsky-Winternitz terms and the related duality transformation are unexplored subjects. The paper is also devoted to the study these systems using representation theory of quadratic algebra developped in context of 2D superintegrable systems.

Let us present the organization of the paper. In Section 2, we recall results obtained in context of 2D systems concerning the realizations of quadratic algebras as deformed oscillator algebras and the representation theory. In Section 3, we recall results concerning the five-dimensional Kepler system and introduce a generalized 5D system. We obtain its integrals of motion and we show that they generate a quadratic algebra with structure constants involving an integral that commute with the generators of this quadratic algebra. This integral is also the Casimir of a $so(4)$ Lie algebra. We present an algebraic derivation of the energy spectrum of this system. In Section 4, we introduce a generalized Yang-Coulomb monopole, present its integrals of motion, the energy spectrum and the wavefunctions using parabolic coordinates. We will show also that this system is related to a 8D singular oscillator by a Hurwitz transformation. We will study the quadratic algebra of this system and present the finite dimensional unitary representations. This quadratic algebra involve two integrals that commute with the generators that are the Casimir of $so(4)$ Lie algebras. We also use these results to reobtain the energy spectrum of the generalized Yang-Coulomb monopole algebraically.

\section{Quadratic algebras}
Quadratic and more generally polynomial algebras were introduced in the context of non-relativistic quantum mechanics [22,23,55-59] 20 years ago. Their finite-dimensional unitary representations can be used to calculate algebraically the energy spectrum of superintegrable systems. Let us recall results obtained in context of two-dimensional quantum superintegrable systems. We refer the reader to the Ref.60 for more detailed discussions. 

In the case of a two-dimensional Hamiltonian $H$ allowing two second order integrals $A$ and $B$ (i.e. $[H,A]=[H,B]=0$), we have the following graph ( as in Ref. 72 )

\begin{equation}
\begin{xy}
(0,0)*+{A}="a"; (20,0)*+{H}="h"; (40,0)*+{B}="b";
"a";"h"**\dir{--}; 
"h";"b"**\dir{--};
\end{xy}
.
\end{equation}

The most general quadratic algebra generated by these integrals is given by the following commutation relations [60] :
\begin{equation}
[A,B]=C,\quad [A,C]=\alpha A^{2} +\gamma\{A,B\}+\delta A+\epsilon B+\zeta, 
\end{equation}
\[ [B,C]=aA^{2}-\gamma B^{2}-\alpha \{A,B\}+d A-\delta B + z.\]
The structure constants $\gamma$, $\delta$, $\alpha$, $\epsilon$, $\zeta$ and $z$ of the quadratic algebra given by Eq.(4) are polynomials of the Hamiltonian $H$. The degree of these polynomials can be deduced from the order of these commutators. The structure constants $\alpha$, $\gamma$ and $a$ are constant and

\begin{equation}
\delta=\delta_{0}+\delta_{1}H,\quad \epsilon=\epsilon_{0}+\epsilon_{1}H,\quad d=d_{0}+d_{1}H ,
\end{equation}
\[ \zeta=\zeta_{0}+\zeta_{1}H+\zeta_{2}H^{2},\quad z=z_{0}+z_{1}H+z_{2}H^{2}. \]

The Casimir operator (i.e. $[K,A]=[K,B]=[K,C]=0$) of this quadratic algebra is thus given in terms of the generators ($A$,$B$ and $C$) by
\begin{equation}
K=C^{2}-\alpha\{A^{2},B\}-\gamma\{A,B^{2}\}+(\alpha \gamma -\delta)\{A,B\}+(\gamma^{2}-\epsilon)B^{2}
\end{equation}
\[+(\gamma \delta -2 \zeta)B+\frac{2a}{3}A^{3}+(d+\frac{a\gamma}{3}+\alpha^{2})A^{2}+(\frac{a\epsilon}{3}+\alpha\delta+2z)A.\]

In order to obtain the finite-dimensional unitary representations, we consider realizations of the quadratic algebras in terms of deformed oscillators algebras [56] $\{N,b^{\dagger},b\}$ satisfying the following equations :
\begin{equation}
[N,b^{\dagger}]=b^{\dagger},\quad [N,b]=-b,\quad bb^{\dagger}=\Phi(N+1),\quad b^{\dagger}b=\Phi(N), 
\end{equation}
where $\Phi(x)$ is a real function called the structure function satisfying $\Phi(0)=0$ and $\Phi(x)>0$ for $x>0$. We have the existence of finite dimensional unitary representations when we impose the existence of a positive integer $p$ such that $\Phi(p+1)=0$. The realizations of the quadratic algebra given by Eq.(4) of the form ($A=A(N)$, $B=b(N)+b^{\dagger}\rho(N)+\rho(N)b$) were studied [60]. 
Two cases were obtained (i.e. $\gamma \neq 0$ and $\gamma = 0$). Let us only present the results in the case $\gamma \neq 0$ with the following choice of structure constants ($\alpha=a=\delta=0$) similarly to the generalized MICZ-Kepler systems ( $u$ is a constant determined from the constraints on the structure function ):

\begin{equation}
\rho(N)=\frac{1}{2^{12}3\gamma^{8}(N+u)(1+N+u)(1+2(N+u))^{2}}
\end{equation}
\begin{equation*}
A(N)=\frac{\gamma}{2}((N+u)^{2}-\frac{1}{4}-\frac{\epsilon}{\gamma^{2}},\quad b(N)=-\frac{\zeta}{\gamma^{2}}\frac{1}{((N+u)^{2}-\frac{1}{4})},
\end{equation*}
\begin{equation*}
\Phi(N)=-3072\gamma^{6}K(-1+2(N+u))^{2}+48d\gamma^{8}(-3+2(N+u))(-1+2(N+u))^{4}(1+2(N+u))\label{eq12}
\end{equation*}
\[+12288\gamma^{4}\zeta^{2}+32\gamma^{4}(-1+2(N+u))^{2}(-1-12(N+u)+12(N+u)^{2})(-4d\epsilon\gamma^{2}+8\gamma^{3}z)\]
\[-256\gamma^{2}(-1+2(N+u))^{2}(-3d\epsilon^{2}\gamma^{2}+2d\epsilon\gamma^{4}+12\epsilon\gamma^{3}z-4\gamma^{5}z).\]

The Casimir operator $K$ can be written in terms of the Hamiltonian only. We have a energy dependent Fock space of dimension p+1 if
\begin{equation}
\Phi(p+1,u,E)=0, \quad \Phi(0,u,E)=0,\quad \Phi(x)>0, \quad \forall \; x>0 \quad .
\end{equation}

The Fock space is defined by
\begin{equation*}
H|E,n>=E|E,n>,\quad N|E,n>=n|E,n> \quad b|E,0>=0.
\end{equation*}
\begin{equation*}
b^{\dagger}|n>=\sqrt{\Phi(n+1,E)}|E,n+1>,\quad b|n>=\sqrt{\Phi(n,E)}|E,n-1>.
\end{equation*}
The energy $E$ and the constant $u$ are solutions of the equations obtained by the system given by Eq.(9). They represent the finite-dimensional unitary representations with dimension $p+1$.

\subsection{Higher dimensional systems and quadratic algebras}

These results obtained in context of two-dimensional systems can be applied to various systems in three and four dimensions.
The Kepler system [55], the harmonic oscillator [56] and the MICZ-Kepler systems on three sphere [73,31], the Hartman potential [23], the  generalized Kepler [22], generalized MICZ-Kepler systems [36], the 4D singular oscillator [36], the generalized Kaluza-Klein monopole [70] and the non-degenerate extended Kepler-Coulomb system [72] possess a quadratic algebra with three generators.

We allow in context of higher dimensional superintegrable system the structure constants $\alpha$, $\delta$, $\gamma$, $\epsilon$, $\zeta$, $a$, $d$ and $z$ to be polynomial functions not only of the Hamiltonian but also of any other integrals of motion ($F_{i}$) that commute with the Hamiltonian, with each other and also with the generators of the quadratic algebra $A$, $B$ and $C$ (that is, $[F_{i},H]=[F_{i},F_{j}]=[F_{i},A]=[F_{i},B]=[F_{i},C]=0$). 

\begin{equation}
\begin{xy}
(0,0)*+{A}="a"; (20,0)*+{F_{i}}="f"; (40,0)*+{B}="b"; (20,20)*+{H}="h"; (60,0)*+{F_{i}}="c"; (80,0)*+{F_{j}}="d";
"a";"h"**\dir{--}; 
"h";"b"**\dir{--};
"h";"f"**\dir{--};
"f";"b"**\dir{--};
"f";"a"**\dir{--};
"c";"d"**\dir{--};
\end{xy}
\end{equation}

When we study this algebra's realization in terms of deformed oscillator algebras and its representations, we will fixed the energy ($H\psi=E\psi$) but also these other integrals ($F_{i}\psi=f_{i}\psi$) of motion that form, with the Hamiltonian, an Abelian subalgebra. The Casimir operator $K$ of this quadratic algebra is thus given in terms of the generators and will be also rewritten as a polynomial of $H$ and $F_{i}$. 

\section{Generalized five-dimensional Kepler}
\subsection{Five-dimensional Kepler system}
Let us first recall some known results concerning the 5D-Kepler system given by :
\begin{equation}
H=\frac{1}{2}P^{2}-\frac{c_{0}}{r},
\end{equation}
with $r=\sqrt{x_{0}^{2}+x_{1}^{2}+x_{2}^{2}+x_{3}^{2}+x_{4}^{2}}$ and
\begin{equation}
P_{i}=-i\hbar\partial_{i},\quad [x_{i},P_{j}]=i\hbar \delta_{ij}.
\end{equation}
This system has the following integrals of motion ($[H,L_{ij}]=[H,M_{k}]=0$)
\begin{equation}
L_{ij}=x_{i}P_{j}-x_{j}P_{i},\quad M_{k}=\frac{1}{2}(P_{i}L_{ik}+L_{ik}P_{i})+\frac{c_{0} x_{k}}{r}.
\end{equation}
These integrals generate the following dynamical symmetry algebra :
\begin{equation}
[L_{ij},L_{mn}]=i\hbar \delta_{im}L_{jn}-i\hbar \delta_{jm}L_{in}-i\hbar \delta_{in}L_{jm}+i\hbar \delta_{jn} L_{im},
\end{equation}
\begin{equation}
[L_{ij},M_{k}]=i\hbar\delta_{ik}M_{j}-i\hbar\delta_{jk}M_{i},\quad [M_{i},M_{k}]=-2i\hbar H L_{ik}.
\end{equation}
The relations given by Eq.(14) generate a $so(5)$ Lie algebra. The algebra generated when we consider also the Eq.(15) and we fixed the energy is isomorphic 
to the $so(6)$ Lie algebra for the bound states. The symmetry algebra can also be seen as a Kac-Moody algebra [79]. This dynamical symmetry algebra can be used 
to obtain algebraically the energy spectrum.

\subsection{Generalized 5D Kepler system}

Let us now introduce the following Hamiltonian :
\begin{equation}
H=\frac{1}{2}P^{2}-\frac{c_{0}}{r}+\frac{c_{1}}{r(r+x_{0})}+\frac{c_{2}}{r(r-x_{0})}.
\end{equation}
The two last terms break the $so(5)$ rotational invariance. This systems is superintegrable and has the following integrals of motion :
\begin{equation}
A= L^{2}+\frac{2rc_{1}}{(r+x_{0})}+\frac{2rc_{2}}{(r-x_{0})},\quad B= M_{0}+\frac{c_{1}(r-x_{0})}{r(r+x_{0})}-\frac{c_{2}(r+x_{0})}{r(r-x_{0})},
\end{equation}
\begin{equation}
L^{2}=\sum_{1<i<j<4}L_{ij}^{2},\quad L_{ij}=x_{i}P_{j}-x_{j}P_{i},
\end{equation}
where i,j=1,2,3,4.

We have the commutation relations :
\begin{equation}
[H,A]=[H,B]=[H,L^{2}]=[A,L^{2}]=[B,L^{2}]=[H,L_{ij}]=0,
\end{equation}
and in terms of graph :

\begin{equation}
\begin{xy}
(0,0)*+{A}="a"; (20,0)*+{L^{2}}="f"; (40,0)*+{B}="b"; (20,20)*+{H}="h";  (60,0)*+{L^{2}}="j"; (100,0)*+{L_{ij}}="l"; (80,20)*+{H}="g"; 
"a";"h"**\dir{--}; 
"h";"b"**\dir{--};
"h";"f"**\dir{--};
"f";"b"**\dir{--};
"f";"a"**\dir{--};
"j";"l"**\dir{--}; 
"l";"g"**\dir{--};
"g";"j"**\dir{--};
\end{xy}
\end{equation}

These integrals generate the following quadratic algebra 

\begin{equation}
[A,B]=C,\quad [A,C]=2\hbar^{2}\{A,B\}+8\hbar^{4}B-4(c_{1}-c_{2})\hbar^{2}c_{0},
\end{equation}
\begin{equation}
[B,C]=-2\hbar^{2}B^{2}+8\hbar^{2}HA-4\hbar^{2}L^{2}H+16\hbar^{4}H-8\hbar^{2}(c_{1}+c_{2})H+2\hbar^{2}c_{0}^{2},
\end{equation}
\begin{equation}
K=16\hbar^{4}HL^{2}-8\hbar^{2}(c_{1}-c_{2})^{2}H+32(c_{1}+c_{2})\hbar^{4}H-32\hbar^{6} H
\end{equation}
\[ +4\hbar^{2}c_{0}^{2}L^{2}+8\hbar^{2}(c_{1}+c_{2})c_{0}^{2}-4\hbar^{4}c_{0}^{2}.\]

This quadratic algebra has a very interesting structure. The structure constants involve the operator $L^{2}$ that is the Casimir operator of the $so(4)$ Lie algebra generated by $L_{ij}$ for $i,j=1,2,3,4$. The $l$ ( $L^{2}\psi=l\psi$) can be calculated from the representation theory and Casimir operators of $so(4)$ Lie algebras.

\subsection{Finite dimensional unitary representations}
We obtain for the generalized 5D Kepler system the following structure function :
\begin{equation*}
\Phi(x)=6191456 E \hbar^{18}(x+u-(\frac{1}{2}(1-m_{1}-m_{2})))(x+u-(\frac{1}{2}(1-m_{1}+m_{2})))(x+u-(\frac{1}{2}(1+m_{1}-m_{2})))
\end{equation*}
\begin{equation}
(x+u-(\frac{1}{2}(1+m_{1}+m_{2})))(x+u-(\frac{1}{2}-\frac{c_{0}}{\sqrt{-2E}\hbar}))(x+u-(\frac{1}{2}+\frac{c_{0}}{\sqrt{-2E}\hbar}))
\end{equation}
with $\hbar^{2}m_{1,2}^{2}=16c_{1,2}+4l+4\hbar^{2}$. From the constraints on the structure function given by the equation (9) we obtain  $u=\frac{1}{2}+\frac{c_{0}}{\sqrt{-2E}\hbar}$ and
\begin{equation}
E=\frac{-c_{0}^{2}}{\hbar^{2}(p+1+\frac{m_{1}+m_{2}}{2})^{2}},\quad p=0,1,2,3,...
\end{equation}
\begin{equation}
\Phi(x)=\frac{6291456\hbar^{18}x(p+1-x)c_{0}^{2}}{\hbar^{2}(p+1+m_{1}+m_{2})}(p+1+m_{1}-x)(p+1+m_{2}-x)(p+1+m_{1}+m_{2}). 
\end{equation}

\section{Generalized Yang-Coulomb monopole system and duality}

\subsection{Yang-Coulomb monopole}
The Yang-Coulomb monopole is given by the following Hamiltonian that consists in a 5D-Kepler systems with a $su(2)$ monopole interaction :
\begin{equation}
H=\frac{1}{2}(\pi_{j})^{2}+\frac{\hbar^{2}}{2 r^{2}}T^{2}-\frac{c_{0}}{r},\quad A_{i}^{a}=\frac{2i}{r(r+x0)}\tau_{ij}^{a}x_{j},
\end{equation}
 with $\pi_{j}=-i\hbar \partial_{x_{j}}-\hbar A_{j}^{a}T_{a}$ and
\begin{equation}
T_{1}=i(cos(\alpha_{T})cot(\beta_{T})\frac{\partial}{\partial \alpha_{T}}+sin(\alpha_{T})\frac{\partial}{\partial \beta_{T}}-\frac{cos(\alpha_{T})}{sin(\beta_{T})}\frac{\partial}{\partial \gamma_{T}}),
\end{equation}
\begin{equation}
T_{2}=i(sin(\alpha_{T})cot(\beta_{T})\frac{\partial}{\partial \alpha_{T}}-cos(\alpha_{T})\frac{\partial}{\partial \beta_{T}}-\frac{sin(\alpha_{T})}{sin(\beta_{T})}\frac{\partial}{\partial \gamma_{T}}),
\end{equation}
\begin{equation}
T_{3}=-i\frac{\partial}{\partial \alpha_{T}},
\end{equation}
with (i,j=0,1,2,3,4 ; $\mu$,$\nu$=1,2,3,4).These operators given by Eq.(27)-(30) satisfy the following relations :
\begin{equation}
[T_{a},T_{b}]=i\epsilon_{abc}T_{c}.
\end{equation}

We have :
\begin{equation} \tau^{1}=\frac{1}{2} \begin{pmatrix} 0 & 0 & 0\\ 0 & 0 & -i\sigma^{1}\\ 0 & i\sigma^{1} & 0  
\end{pmatrix},\quad \tau^{2}=\frac{1}{2} \begin{pmatrix} 0 & 0 & 0\\ 0 & 0 & i\sigma^{3}\\ 0 & -i\sigma^{3} & 0  \end{pmatrix},\quad \tau^{3}=\frac{1}{2} \begin{pmatrix} 0 & 0 & 0\\ 0 & \sigma^{2} & 0\\ 0 &0 & \sigma^{2}  \end{pmatrix},
\end{equation}
where $\sigma^{i}$ for $i=1,2,3$ are the Pauli matrices :
\begin{equation}
\sigma^{1}= \begin{pmatrix} 0 & 1   \\ 1 & 0  \end{pmatrix},\quad \sigma^{2}=\begin{pmatrix} 0 & -i   \\ i & 0   \end{pmatrix},\quad \sigma^{3}= \begin{pmatrix} 1 & 0  \\0 & -1  \end{pmatrix}.
\end{equation}

We can construct the following field tensor

\begin{equation}
F_{ik}^{a}=\partial_{i}A_{k}^{a}-\partial_{k}A_{i}^{a}+\epsilon_{abc}A_{i}^{b}A_{k}^{c},
\end{equation}

and using Eq(28)-(30) and (34) generate the following integrals of motion :

\begin{equation}
L_{ik}=(x_{i}\pi_{k}-x_{k}\pi_{i})-r^{2}\hbar F_{ik}^{a}T_{a},\quad M_{k}=\frac{1}{2}(\pi_{i}L_{ik}+L_{ik}\pi_{i}+2 c_{0}\frac{x_{k}}{r})
\end{equation}

They satisfy Eq.(14) and (15) and thus preserve the $so(6)$ Lie algebra.

\subsection{Yang-Coulomb monopole with Smorodinsky-Winternitz terms}
We can introduced the following Hamiltonian :

\begin{equation}
H=\frac{1}{2}(\pi_{j})^{2}+\frac{\hbar^{2}}{2 r^{2}}T^{2}-\frac{c_{0}}{r}+\frac{c_{1}}{r(r+x_{0})}+\frac{c_{2}}{r(r-x_{0})}.
\end{equation}
Similarly, we can show with more involving calculations that this system allows integrals of the form given by equation (17) and (18) where the $L_{ij}$ and $M_{0}$ are given by Eq.(35). This system is thus a new superintegrable system and multiseparable ( a consequence of this existence of more than one second order integrals of motion). It allows separation of variables  in hyperspherical, spheroidal and parabolic coordinates systems. Let us present the wavefunctions and a derivation of the energy spectrum of this systems using the parabolic coordinates. We define the parabolic coordinates by :
\begin{equation}
x_{0}=\frac{1}{2}(\mu-\nu),
\end{equation}
\[x_{1}+ix_{2}=\sqrt{\mu\nu}cos(\frac{\beta}{2})e^{\frac{i(\alpha+\gamma)}{2}},\]
\[x_{3}+ix_{4}=\sqrt{\mu\nu}sin(\frac{\beta}{2})e^{\frac{i(\alpha-\gamma)}{2}},\]

with $\nu, \mu \in [0,\infty )$. The Schroedinger equation given given by the Hamiltonian (36) (with $H\psi=\epsilon\psi$) become (with  $J_{i}=L_{i}+T_{i}$)

\begin{equation}
(\Delta_{\mu,\nu} -\frac{(4J^{2}-2c_{1})}{\mu(\mu+\nu)}- \frac{(4L^{2}-2c_{2})}{\nu(\mu+\nu)})\psi +  \frac{2}{\hbar^{2}}(\epsilon+\frac{2c_{0}}{\mu+\nu})\psi=0 ,
\end{equation}

with

\begin{equation}
\Delta_{\mu,\nu}=\frac{4}{\mu+\nu}(  \frac{1}{\mu}\frac{\partial}{\partial \mu}(\mu^{2}\frac{\partial}{\partial \mu})+ \frac{1}{\nu}\frac{\partial}{\partial \nu}(\nu^{2}\frac{\partial}{\partial \nu})).
\end{equation}

We use the following form for the wavefunctions (similarly to the Yang-Coulomb monopole [44])
\begin{equation}
\psi=C f_{1}(\mu)f_{2}(\nu) D_{L T m' t'}^{JM}(\alpha,\beta,\gamma,\alpha_{T},\beta_{T},\gamma_{T}),
\end{equation}

with $D_{L T m' t'}^{JM}(\alpha,\beta,\gamma,\alpha_{T},\beta_{T},\gamma_{T})$ given by

\begin{equation}
D_{L T m' t'}^{JM}(\alpha,\beta,\gamma,\alpha_{T},\beta_{T},\gamma_{T})=\sqrt{\frac{(2L+1)(2T+1)}{4\pi^{4}}}\sum_{M=m+t}C_{L,m;T,t}^{JM}D_{mm'}^{L}(\alpha,\beta,\gamma)D_{tt'}^{T}(\alpha_{T},\beta_{T},\gamma_{T}),
\end{equation}

where $C_{L,m;T,t}^{JM}$ are the Clebseh-Gordon coefficients, $D_{mm'}^{L}(\alpha,\beta,\gamma)$ and $D_{tt'}^{T}(\alpha_{T},\beta_{T},\gamma_{T})$ are the Wigner functions, and C is a normalization constant.
Using the following parameters $s_{1}=4(j(j+1)-2c_{1})$, $s_{2}=4(l(l+1)-2c_{2})$, $\beta=\frac{2\epsilon}{\hbar^{2}}$ and  $\alpha=\frac{2c_{0}}{\hbar^{2}}$ and the wavefunctions as given by Eq.(40), we obtain two seconder order differential equations :
\begin{equation}
(\partial_{\mu}(\mu \partial_{\mu})-\frac{s_{1}}{4\mu}+\frac{\alpha}{4}+\frac{\beta}{4}\mu)f_{1}(\mu)=\frac{v}{2}f_{1}(\mu),
\end{equation}

\begin{equation}
(\partial_{\nu}(\nu \partial_{\nu})-\frac{s_{2}}{4\nu}+\frac{\alpha}{4}+\frac{\beta}{4}\nu)f_{2}(\nu)=-\frac{v}{2}f_{2}(\nu).
\end{equation}

The solution can be written in term of the hypergeometric function (with $x=\mu \sqrt{-\beta}$ and $y=\nu \sqrt{-\beta}$)
\begin{equation}
f_{1}(\mu)=x^{\frac{s_{1}}{2}}e^{-\frac{x}{2}}F(-n_{1},s_{1}+1,x), \quad f_{2}(\nu)=y^{\frac{s_{2}}{2}}e^{-\frac{y}{2}}F(-n_{2},s_{2}+1,y), 
\end{equation}

where  $n_{1},n_{2}=0,1,2,...$ and

\begin{equation}
n_{1}=-\frac{s_{1}+1}{2}+\frac{(2v-\alpha)}{4\sqrt{-\beta}},\quad n_{2}=-\frac{s_{2}+1}{2}-\frac{(2v+\alpha)}{4\sqrt{-\beta}}.
\end{equation}

From Eq.(45) we have the relation

\begin{equation}
n_{1}+n_{2}+(s_{1}+s_{2}+1)=-\frac{\alpha}{2\sqrt{-\beta}},
\end{equation}

that allows us to obtain the energy spectrum

\begin{equation}
\epsilon=\frac{-c_{0}^{2}}{2\hbar^{2}(n_{1}+n_{2}+(s_{1}+s_{2}+1))^{2}}.
\end{equation}

\subsection{Hurwitz transformation and 8D singular oscillator}

The Schroedinger equation of the generalized Yang-Coulomb monopole system

\begin{equation}
\frac{1}{2}(-i\hbar \frac{\partial}{\partial x_{j}}-\hbar A^{a}_{j} T_{a})^{2}\psi +\frac{\hbar^{2}}{2 r^{2}} T^{2} \psi  - \frac{c_{0}}{r} \psi  +\frac{c_{1}}{r(r+x_{0})}+\frac{c_{2}}{r(r-x_{0})}=\epsilon \psi,
\end{equation}

is related to the following 8D singular oscillator 

\begin{equation}
-\frac{\hbar^{2}}{2}\frac{\partial^{2}}{\partial u^{2}}+\frac{\omega^{2}}{2}u^{2} +\frac{\lambda_{1}}{u_{0}^{2}+u_{1}^{2}+u_{2}^{2}+u_{3}^{2}}+\frac{\lambda_{2}}{u_{4}^{2}+u_{5}^{2}+u_{6}^{2}+u_{7}^{2}}=E,
\end{equation}

by the following Hurwitz transformation (that convert $\mathbb{R}^{8}$ to the direct product  $\mathbb{R}^{5}\otimes \mathbb{S}^{3}$) :
\begin{equation}
x_{0}=u_{0}^{2}+u_{1}^{2}+u_{3}^{2}-u_{4}^{2}-u_{5}^{2},
\end{equation}
\[ x_{1}=2(u_{0}u_{4}-u_{1}u_{5}-u_{2}u_{6}-u_{3}u_{7}),\]
\[ x_{2}=2(u_{0}u_{5}+u_{1}u_{4}-u_{2}u_{7}+u_{3}u_{6}),\]
\[ x_{3}=2(u_{0}u_{6}+u_{1}u_{7}+u_{2}u_{4}-u_{3}u_{5}),\] 
\[ x_{4}=2(u_{0}u_{7}-u_{1}u_{6}+u_{2}u_{5}+u_{3}u_{4}),\]
\[\alpha_{T}=\frac{i}{2}\ln{\frac{(u_{0}+iu_{1})(u_{2}-iu_{3})}{(u_{0}-iu_{1})(u_{2}+iu_{3})}}, \in [0,2\pi) \]
\[\beta_{T}=2\arctan{\frac{u_{0}^{2}+u_{1}^{2}}{u_{2}^{2}+u_{3}^{2}}}^{\frac{1}{2}}, \in [0,\pi]  \]
\[\gamma_{T}=\frac{i}{2}\ln{\frac{(u_{0}-iu_{1})(u_{2}-iu_{3})}{(u_{0}+iu_{1})(u_{2}+iu_{3})}}, \in [0,4\pi) \]

with the following relations between the parameters
\begin{equation}
c_{0}=\frac{E}{4},\quad \epsilon=\frac{-\omega^{2}}{8},\quad c_{i}=2\lambda_{i}.
\end{equation}

The 8D singular oscillator is also superintegrable and has the following integrals of motion

\begin{equation}
J_{ij}=u_{i}\partial_{u_{j}}-u_{j}\partial_{u_{i}},\quad i<j,\quad i,j=0,1,2,3
\end{equation}
\begin{equation}
K_{ij}=u_{i}\partial_{u_{j}}-u_{j}\partial_{u_{i}},\quad i<j,\quad i,j=4,5,6,7
\end{equation}
\begin{equation}
J^{2}=\sum_{1<i<j<4}J_{ij}^{2},\quad K^{2}=\sum_{1<i<j<4}L_{ij}^{2},
\end{equation}
\begin{equation}
A=-\frac{1}{4}(u^{2}\frac{\partial^{2}}{\partial u^{2}}-u_{i}u_{j}\frac{\partial^{2}}{\partial_{u_{i}}\partial_{u_{j}}}-7u_{i}\frac{\partial}{\partial u_{i}})
\end{equation}
\[+(\frac{1}{2\hbar^{2}}u^{2}(   +\frac{\lambda_{1}}{u_{0}^{2}+u_{1}^{2}+u_{2}^{2}+u_{3}^{2}}+\frac{\lambda_{2}}{u_{4}^{2}+u_{5}^{2}+u_{6}^{2}+u_{7}^{2}}),\]
\begin{equation}
B=\frac{\hbar^{2}}{2}(\frac{\partial^{2}}{\partial u^{2}}+\frac{\omega^{2}}{2}(u_{0}^{2}+u_{1}^{2}+u_{2}^{2}+u_{3}^{2}-u_{4}^{2}-u_{5}^{2}-u_{6}^{2}-u_{7}^{2})+\frac{\lambda_{1}}{u_{0}^{2}+u_{1}^{2}+u_{2}^{2}+u_{3}^{2}}-\frac{\lambda_{2}}{u_{4}^{2}+u_{5}^{2}+u_{6}^{2}+u_{7}^{2}}.
\end{equation}

We have the following commutation relations :
\begin{equation}
[H,A]=[H,B]=[H,K^{2}]=[H,J^{2}]=[H,J_{ij}]=[H,K_{ij}]=0
\end{equation}
\[ [A,J^{2}]=[A,K^{2}]=[B,J^{2}]=[B,K^{2}]=0\]

and in terms of graph :

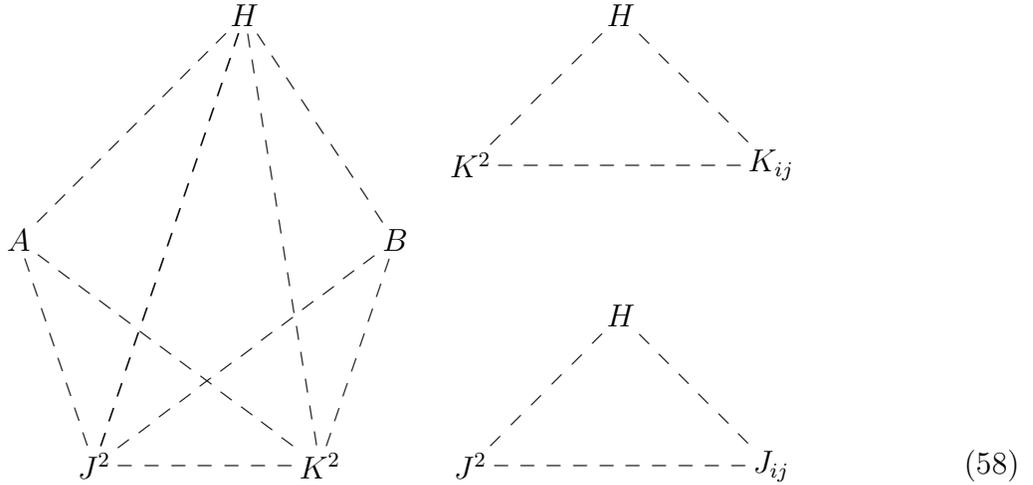
\begin{equation}
\begin{xy}
(10,0)*+{J^{2}}="f"; (40,0)*+{K^{2}}="k"; (0,30)*+{A}="a"; (50,30)*+{B}="b"; (30,60)*+{H}="h";  (60,0)*+{J^{2}}="j1"; (100,0)*+{J_{ij}}="j2"; (80,20)*+{H}="j3";  (60,40)*+{K^{2}}="k1"; (100,40)*+{K_{ij}}="k2"; (80,60)*+{H}="k3"; 
"f";"k"**\dir{--}; 
"h";"b"**\dir{--};
"h";"f"**\dir{--};
"h";"a"**\dir{--};
"h";"k"**\dir{--};
"h";"f"**\dir{--};
"a";"f"**\dir{--}; 
"a";"k"**\dir{--};
"b";"f"**\dir{--};
"b";"k"**\dir{--};
"j1";"j2"**\dir{--};
"j1";"j3"**\dir{--};
"j2";"j3"**\dir{--};
"k1";"k2"**\dir{--};
"k1";"k3"**\dir{--};
"k2";"k3"**\dir{--};
\end{xy}
\end{equation}

The integrals generate the following quadratic algebra

\begin{equation}
[A,B]=C,\quad [A,C]=2\{A,B\}+8B+ J^{2}H-K^{2}H-\frac{2(\lambda_{1}-\lambda_{2})}{\hbar^{2}}H,
\end{equation}
\begin{equation}
[B,C]=4\hbar^{2}B^{2}+2H^{2}-16\hbar^{2}\omega^{2}A-4\hbar^{2}\omega^{2}J^{2}-4\hbar^{2}\omega^{2}K^{2}+8(\lambda_{1}+\lambda_{2}-4\hbar^{2})\omega^{2},
\end{equation}
\begin{equation}
K=-2J^{2}H^{2}-2K^{2}H^{2}+\frac{4(\lambda_{1}+\lambda_{2}-\hbar^{2})\omega^{2}}{\hbar^{2}}H^{2}+\hbar^{2}\omega^{2}(J^{2})^{2}-\hbar^{2}\omega^{2}(K^{2})^{2}-2\hbar^{2}\omega^{2}J^{2}K^{2}
\end{equation}
\[-4(\lambda_{1}-\lambda_{2}-4\hbar^{2})\omega^{2}J^{2}+4(\lambda_{1}-\lambda_{2}+4\hbar^{2})\omega^{2}K^{2}+ \frac{4((\lambda_{1}-\lambda_{2})^{2}-8(\lambda_{1}+\lambda_{2})\hbar^{2}+16\hbar^{4})\omega^{2}}{\hbar^{2}}.\]

The structure constants and the Casimir operator allow us to obtain the structure function $\Phi(x)$ of the deformed oscillator algebra. The energy and the two Casimir operators are fixed, $H\psi=E\psi$, $J^{2}\psi=j\psi$ and $K^{2}\psi=k\psi$. The structure function is then given by (with $\hbar^{2}m_{1}=\hbar^{2}j^{2}+2\lambda_{1}+\hbar^{2}$ and $\hbar^{2} m_{2}=\hbar^{2}k^{2}+2\lambda_{2}+\hbar^{2}$)

\begin{equation}
\Phi(x)=3\cdot 2^{22}\omega^{2}\left(x+u-\left(\frac{1}{2}-\frac{E}{2\omega\hbar}\right)\right)\left(x+u-\left(\frac{1}{2}+\frac{E}{2\omega\hbar}\right)\right)
\end{equation}
\[ \left(x+u-\left(\frac{1}{2}-\frac{1}{2}(m_{1}+m_{2})  \right)\right) \left(x+u-\left(\frac{1}{2}-\frac{1}{2}(m_{1}-m_{2})  \right)\right)\]
\[ \left(x+u-\left(\frac{1}{2}-\frac{1}{2}(-m_{1}+m_{2})  \right)\right) \left(x+u-\left(\frac{1}{2}-\frac{1}{2}(-m_{1}-m_{2})  \right)\right).\]

We thus found the following energy spectrum with the corresponding finite dimensional unitary representation:

\begin{equation}
E=2\omega\hbar(p+1+\frac{m_{1}+m_{2}}{2}),\quad u=\frac{1}{2}-\frac{E}{2\omega\hbar},
\end{equation}
\begin{equation}
\phi(x)=3\cdot 2^{19}\omega^{2}x(p+1-x)(p+1+m_{1}-x)(p+1+m_{2}-x)
\end{equation}
\[(p+1+m_{1}+m_{2}-x)(2p+2+m_{1}+m_{2}-x).\]

Using the relations between the parameter of the generalized Yang-Coulomb monopole and the 8D singular oscillator, we obtain the following equation (with Eq(51))

\begin{equation}
4c_{0}=E=2\omega\hbar(p+1+\frac{m_{1}+m_{2}}{2})=2\sqrt{-8\epsilon}\hbar(p+1+\frac{m_{1}+m_{2}}{2}),
\end{equation}

that can be used to reobtain algebraically the energy spectrum of the generalized Yang-Coulomb monopole
\begin{equation}
\epsilon=\frac{-c_{0}^{2}}{2\hbar^{2}(p+1+\frac{m_{1}+m_{2}}{2})^{2}}.
\end{equation}

This results coincide with the one obtained from a direct approach using the parabolic coordinates and given by Eq.(47) ( with $m_{i}=2s_{i}$ and $p=n_{1}+n_{2}$ ).

\section{Conclusion}

One main results of this paper is the study of three new superintegrable systems the generalized 5D Kepler system, the generalized Yang-Coulomb monopole and the 8D singular oscillator. The study of systems with nonAbelian monopole is important, and this paper point out how they could be generalized using Smorodinsky-Winternitz terms. We show that for the generalized 5D Kepler systems (as for the well studied 5D Kepler systems) the superintegrability property is preserved when one add the appropriate nonAbelian monopole. The Hurwitz transformation maps also this superintegrable system to an other one a 8D singular oscillator.

The algebraic structure presented in this paper are interesting. In the case of the generalized Kepler system the quadratic algebra involve the Casimir operator of a $so(4)$ Lie algebra. In the case of the dual of the generalized Yang-Coulomb monopole, the  8D singular oscillator, the quadratic algebra involve two integrals that are Casimir operator of $so(4)$ Lie algebras. This point out also how the quadratic algebra approach can be apply in context of higher dimensional superintegrable.

Let us mention that these results can have a wider applications, it was shown that systems with monopole can be extended in N dimensions [80] and a generalized Hurwitz transformation were obtained for a 9-dimensional hydrogenlike system with $so(8)$ non-Abelian monopole [81]. Let us mention also recent works on monopoles [82-86].
 
\textbf{Acknowledgments}

The research of I.M. was supported by a postdoctoral research fellowship from FQRNT of Quebec. The author thanks N.Mackay for very helpful comments and discussions. 

\section{\textbf{References}}

[1] W.Pauli, Z.Phys. 36 336 (1926)
\newline
[2] V.Fock, Z.Phys. 98 145 (1935)
\newline
[3] V.Bargmann, Z.Phys. 99 576 (1936)
\newline
[4] J.Hermann, Giornale de Letterati D'Italia, (Appresso G. G. Hertz, Venezia), 2 447 (1710)
\newline
[5] J.Hermann, Histoire de L'académie Royale des Sciences, (Charles-Estienne Hoschereau, Paris) 519 (1710)
\newline
[6] J.I.Bernoulli, Histoire de L'académie Royale des Sciences, (Charles-Estienne Hoschereau, Paris) 521 (1710)
\newline
[7] P.S.Laplace, Trait\'e de m\'ecanique c\'eleste, (Duprat, Paris) (1799)
\newline
[8] P.Winternitz, Ya.A.Smorodinsky, M.Uhlir and I.Fris, Sov. J.Nucl.Phys. 4 444 (1967)
\newline
[9] A.Makarov, Kh.Valiev, Ya.A.Smorodinsky and P.Winternitz,  Nuovo Cim. A 52 1061 (1967)
\newline
[10] N.W.Evans, Phys.Rev. A 41 5666 (1990)
\newline
[11] N.W.Evans, J.Math.Phys. 32 3369 (1991)
\newline
[12] I.Sokmen, Phys.Lett. A 115 249 (1986)
\newline
[13] L.Chetouani, L.Gurchi and T.F.Hammann, Phys. Lett. A 125 277 (1987)
\newline
[14] M.V.Carpio-Bernido, J.Phys.A:Math.Gen 24 3013 (1991) 
\newline
[15] M.V.Carpio-Bernido, C.C.Bernido and A.Inomata, Third International  Conference on Path Integrals for mev to MeV, 442 Eds V.Sa-Yakanit et al (World Scientific, Singapore) (1989)
\newline
[16] M.Kibler, L.G.Mardoyan and G.Pogosyan, Int J Quantum Chem. 52 1301 (1994) 
\newline
[17] M.Kibler and C.Campigotto, Phys.lett. A 181 1 (1993)
\newline
[18] M.Kibler, G.H.Lamot and P.Winternitz, Int J. Quantum Chem. 43 625 (1992)
\newline
[19] F.Calogero, J.Math.Phys. 10 2191 (1969)
\newline
[20] M.Kibler and T.Negadi, Int.J.Quantum chem 26 405 (1984)
\newline
[21] A.Guha and S.Muherjee, J.Math.Phys. 28 840 (1987)
\newline
[22] A.S.Zhedanov, J.Phys.A.:Math.gen. 26 4633 (1993)
\newline
[23] Ya.A.Granovsky, A.S.Zhedanov and I.M.Lutzenko, J.Phys.A:Math.Gen. 4 3887 (1991)
\newline
[24] H.Hartmann, Theor. Chim. Acta. 24 201 (1972)
\newline
[25] H.V.McIntosh and A.Cisneros, J.Math.Phys. 11 896 (1970)
\newline
[26] D.Zwanziger, Phys. Rev. 167 1480 (1968)
\newline
[27] R.Jackiw, Ann. Phys. 129 183 (1980)
\newline
[28] A.O.Barut, J.Phys. A: Math. Gen. 14 L267 (1981)
\newline
[29] H.Bacry, J.Phys. A: Math. Gen. 14 L73 (1981)
\newline
[30] E.D'Hoker and L.Vinet, Phys. Lett. B 137 72 (1984)
\newline
[31] L.Feher, J.Phys. A: Math. Gen. 19 1259 (1986)
\newline
[32] L.Mardoyan, J.Math.Phys. 44 11 (2003)
\newline
[33] L.G.Mardoyan and M.G.Petrosyan, Phys. Atom. Nucl. 70 572 (2007)
\newline
[34] P.Ranjan Giri, Mod.Phys.Lett. A 23 895 (2008)
\newline
[35] M.Salazar-Ramirez, D.Martinez, V.D.Granados and R.D.Mota, Int.J.Theor.Phys. 49 967 (2010)
\newline
[36] I.Marquette, J.Math.Phys. 51 102105 (2010)
\newline
[37] E.Tresaco and S.Ferrer, Celest.Mech.Dyn.Astr. 107 337 (2010)
\newline
[38] M.Petrosyan, Phys. Atom. Nucl. 71 1094 (2008)
\newline
[39] P.Kustaanheimo and E.Stiefel, J.Reine angew. Math. 218 204 (1965)
\newline
[40] T.Levi-Civita, Opere Mathematische 2 411 (1951)
\newline
[41] A.Hurwitz, Mathematische Werke (Band II Birkhausser, Basel) 641 (1993)
\newline
[42] L.G.Mardoyan, A.N.Sissakian and V.M.Ter-Antonyan, Mod.Phys. Lett. A 14 1303 (1999)
\newline
[43] L.G.Mardoyan, A.N.Sissakian, V.M.Ter-Antonyan, Phys.Atom. Nucl. 61 1746 (1998)
\newline
[44] L.G.Mardoyan, A.N.Sissakian, V.M.Ter-Antonyan, Theor. Math. Phys. 123 1 451 (2000) 
\newline
[45] A.Nersessian and G.Pogosyan, Phys. Rev. A 63 020103 (2001)
\newline
[46] L.G.Mardoyan, Phys. Atom. Nucl. 65 6 1096 (2002)
\newline
[47] M.V.Pletyukhov and E.A.Tolkachev, Report on Math. Phys. 43 1/2 303 (1999)
\newline
[48] E.G.Kalnins, W.Miller Jr and G.S.Pogosyan, J.Math.Phys. 41 5 2629 (2000)
\newline
[49] M.V.Pletyukhov and E.A.Tolkachev, J.Math.phys. 40 1 (1999)
\newline
[50] M.V.Pletyukhov and E.A.Tolkachev, J.Phys.A: Math. Gen. 32 L249 (1999)
\newline
[51] M.Trunk, Int. J.Mod.Phys.A 11 2329 (1996)
\newline
[52] I.Marquette, J. Math. Phys. 50 012101 (2009)
\newline
[53] I.Marquette, J.Math.Phys. 50 095202 (2009)
\newline
[54] I.Marquette, J.Phys.A: Math. Gen. 43 135203  (2010)
\newline
[55] Ya.I.Granovskii, A.S.Zhedanov and I.M.Lutzenko, Theoret. and Math. Phys. 89 474 (1992)
\newline
[56] Ya.I.Granovskii, A.S.Zhedanov and I.M.Lutzenko, Theoret. and Math. Phys. 91 604 (1992)
\newline
[57] D.Bonatsos, C.Daskaloyannis and K.Kokkotas, Phys.Rev. A 48 R3407 (1993)
\newline
[58] D.Bonatsos, C.Daskaloyannis and K.Kokkotas, Phys. Rev. A 50 3700 (1994)
\newline
[59] P.L\'etourneau and L.Vinet, Ann. Phys. 243 144 (1995)
\newline
[60] C.Daskaloyannis, J.Math.Phys. 42 1100 (2001)
\newline
[61] E.G.Kalnins, J.M.Kress, W.Miller Jr and S.Post, SIGMA 5 008 (2009)
\newline
[62] E.G.Kalnins, W.Miller Jr and S.Post, SIGMA 008 (2008)
\newline
[63] C.Daskaloyannis and Y.Tanoudis, Talk XXVII Colloquium on Group Theoretical Methods in Physics, Yerevan, Armenia, Aug, ( arXiv:0902.0130) (2008)
\newline
[64] C.Daskaloyannis and Y.Tanoudis, Contribution to the 4th Workshop on Differential Equations and Group Analysis of Integrable systems, Protaras, Cyprus, Oct (  arXiv:0902.0259 ) (2008)
\newline
[65] C.Daskaloyannis and Y.Tanoudis, Phys.Atom.Nucl. 73 2 214 (2010)
\newline
[66] C.Quesne, SIGMA 3 016 (2007)
\newline
[67] I.Marquette, J.Math.Phys. 51 072903  (2010)
\newline
[68] I.Marquette, J.Phys.:Conf.Series 284 012047 (2011)
\newline
[69] I.Marquette, J.Math.Phys. 52 04230 (2011)
\newline
[70] I.Marquette, J.Phys.A: Math. Gen. 44 235203 (2011)
\newline
[71] C.Daskaloyannis, J.Phys.A: Math.Gen. 24 L789 (1991)
\newline
[72] Y.Tanoudis and C.Daskaloyannis, SIGMA 7 054, 11 pages (2011)
\newline
[73] E.G.Kalnins, W.Miller Jr and S.Post, SIGMA 4 008 (2008) 
\newline
[74] S.Post, SIGMA 7 036 (2011)
\newline
[75] J.M.Kress, Phys. Atomic Nuclei 70 560 (2007) 
\newline
[76] C.Daskaloyannis and Y.Tanoudis, Phys. Atomic Nuclei 71 853 (2008) 
\newline
[77] C.Daskaloyannis and Y.Tanoudis, J.Math.Phys. 48 072108 (2007) 
\newline
[78] V.V.Gritsev, Yu.A.Kurochkin and V.S.Otchik, J.Phys.A: Math. Gen 33 4903 (2000)
\newline
[79] J.Daboul, P.Slodowy and C.Daboul, Phys. Lett. B 317 321 (1993)
\newline
[80] G.Meng G, J.Phys. Math. 48 032105 (2007) 
\newline
[81] V.-H.Le, T.-S.Nguyen and N.H.Phan, J.Phys. A:Math. Theor. 42 175204 (2009) 
\newline
[82] P.A.Horvathy and J.-P.Ngome, Phys. Rev. D 79 127701 (2009)
\newline
[83] J.-P.Ngome, J.Math.Phys. 50 122901 (2009)
\newline
[84] P.M.Zhang, P.A.Horvathy, J.-P.Ngome, Phys Lett A 374 4275 (2010)
\newline
[85] J.-P.Ngome, PhD Thesis, (Super)symmetries of semiclassical models in theoretical and
condensed matter physics, Tours University, 139 pages, ( Preprint arXiv:1103.4876) (2011)
\newline
[86] S.Krivonos, A.Nersessian and V.Ohanyan, Phys. Rev. D {\bf 75} 085002 (2007)

\end{document}